\documentclass[epj,nopacs,final]{svjour}
\usepackage[dvips]{graphicx}
\usepackage{here}
\usepackage[dvips]{epsfig}
\usepackage{bm}
\newcommand{\npb}[3]{{Nucl. Phys. B} {\bf #1}  (#2) #3}
\newcommand{\prd}[3]{{Phys. Rev. D} {\bf #1}  (#2) #3}
\newcommand{\prl}[3]{{Phys. Rev. Lett.} {\bf #1}  (#2) #3}
\newcommand{\plb}[3]{{Phys. Lett. B} {\bf #1}  (#2) #3}
\newcommand{\ibid}[3]{{ibid.} {\bf #1}  (#2) #3}
\newcommand{\zpc}[3]{{Z. Phys. C} {\bf #1}  (#2) #3}
\newcommand{\ijmpa}[3]{{Int. J. Mod. Phys. A} {\bf #1}  (#2) #3}

\newcommand{\ordo}[1]{\ensuremath{{\cal O}(#1)}}
\begin{document}

\title{
Review on the inclusive rare decays 
$\bm{B \to X_s \gamma}$ and
$\bm{B \to X_d \gamma}$ in the Standard Model} 
\author{Kay Bieri
   \and Christoph Greub
}                     
\institute{ 
  Institut f\"ur Theoretische Physik,
  Universit\"at Bern, CH-3012 Bern, Switzerland}
\date{Received: date / Revised version: date}
%
\abstract{We review the NLL QCD calculations for the branching ratio
of $B \to X_s \gamma$ in the SM. In particular, we emphasize the problem
related to the definition of the charm quark mass which leads to a rather
large uncertainty of the NLL predictions. The various steps needed for a
NNLL calculation, in which the $m_c$ issue can be settled, is also sketched.
We briefly summarize the results of a calculation of the 
$\ordo{\alpha_s^2 n_f}$ corrections to ${\rm BR}(B \to X_s \gamma)$, 
which was recently performed as a first step in the NNLL program. We
then also briefly review the status of the photon energy spectrum
and show the comparison with experimental data. Finally, we review
the status of the CKM suppressed decay mode $B \to X_d \gamma$. 
\PACS{
      {PACS-key}{describing text of that key}   \and
      {PACS-key}{describing text of that key}
     } 
} 
\titlerunning{$B \to X_s \gamma$ and $B \to X_d \gamma$ in the SM}
\authorrunning{K. Bieri and C. Greub}
\maketitle
\section{Introduction}
\label{sec:introduction}

In the Standard model (SM), rare $B$ decays like 
$B \to X_s \gamma$ or $B \to X_s \ell^+ \ell^-$ are
induced by one-loop diagrams, where virtual $W$ bosons and
up-type quarks are exchanged. In many extensions of the SM, there
are additional contributions, where the SM particles in the loop
are replaced by nonstandard ones, like charged Higgs bosons, gluinos,
charginos etc. If the masses of these new particles are not 
heavier by many orders of magnitude than the heaviest SM particles,
the new physics contributions to rare $B$ meson decays are expected 
to be generically large. The sensitivity for nonstandard effects 
implies the possibility for an indirect observation of new physics, or 
allows to put limits on the masses and coupling parameters of the new
particles.

It is obvious that it is only possible to fully exploit the new
physics potential of these decays when both, precise
measurements and precise theoretical SM calculations exist.

\begin{figure}[tb]
\centerline{
\resizebox{0.85\hsize}{!}{
\includegraphics*{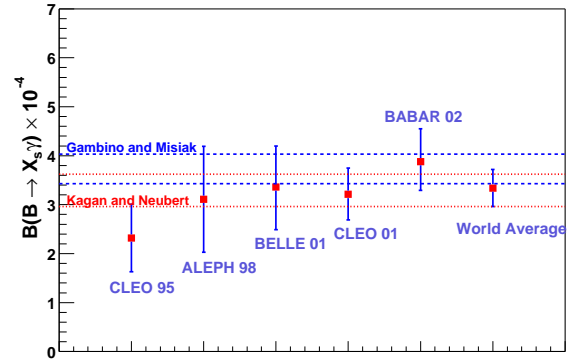}
}}
\caption{Branching ratio ${\rm BR}(B \to X_s \gamma)$: The numbers attached
to the various experiments reflect the year of publication of the 
corresponding result. The dashed (dotted) band shows the theoretical results
based on the MS-bar (pole mass) interpretation of the charm quark mass; see
eqs. (\ref{NLLmcmsbar}) and (\ref{NLLmcpole}). 
Figure taken from \cite{BABAR02} and world average added from
\cite{Jessop2003}.}
\label{fig:1}
\end{figure}

In the following we mainly concentrate on the decays $B \to X_s \gamma$
and $B \to X_d \gamma$, while the rare semileptonic decay $B \to X_s \ell^+
\ell^-$ is reviewed at this conference by T. Hurth \cite{Hurth_conf}. 
There are experimental analyses of the branching ratio 
${\rm BR}(B \to X_s \gamma)$ by 
CLEO \cite{CLEOold,CLEOmed,CLEOnew}, ALEPH \cite{BRALEPH}, 
BELLE \cite{BELLE01}, and BABAR \cite{BABAR02} as shown in fig. \ref{fig:1},
%
%
%
%
leading to the world average \cite{Jessop2003}
\[ {\rm BR}(B \to X_s \gamma)_{\rm{exp}} = (3.34 \pm 0.38) \times 10^{-4}. \]

In contrast to the exclusive rare decay $B \to K^* \gamma$, 
the inclusive counterpart $B \to X_s \gamma$ 
is theoretically much cleaner as no specific model is
needed to describe the hadronic final state. Indeed, nonperturbative effects
in the inclusive decay mode are well under control due to the heavy quark
expansion technique (HQE), which implies that the decay width 
$\Gamma(B \to X_s \gamma)$ is well approximated by the partonic decay rate
$\Gamma(b \to X_s \gamma)$ which can be analyzed in renormalization 
group improved perturbation theory. The (nonperturbative) power corrections
which scale like $1/m_b^2$ \cite{powermb} and $1/m_c^2$ \cite{powermc} 
were estimated to be well below
$10\%$.

\section{Theoretical framework}
Short distance QCD effects 
enhance the partonic decay rate $ \Gamma(b \to s \gamma)$ 
by more than a factor of two.
Analytically, these QCD corrections contain large logarithms of the form 
$\alpha_s^n(m_b) \, \ln^m(m_b/M)$,
where $M=m_t$ or $M=m_W$ and $m \le n$ (with $n=0,1,2,...$).
In order to get a reasonable prediction for 
the decay rate, it turns out that one has to resum both, 
the leading-log (LL) terms ($m=n$) as well as the next-to-leading-log
(NLL) terms ($m=n-1$). 

To achieve the necessary resummations, one usually constructs in a first
step
an effective low-energy theory and then resums the large logarithms
by renomalization group techniques. The low energy theory is
obtained by integrating out the
heavy particles which in the SM are the top quark and the $W$-boson. 
The resulting effective Hamiltonian relevant for $b \to s \gamma$ 
in the SM and many of its extensions reads 
\begin{equation}
\label{heff}
H_{\rm{eff}}(b \to s \gamma)
       = - \frac{4 G_{F}}{\sqrt{2}} \, \lambda_{t} \, \sum_{i=1}^{8}
C_{i}(\mu) \, O_i(\mu) \quad , 
\end{equation}
where $O_i(\mu)$ are local operators consisting of light fields,
$C_{i}(\mu)$ are the corresponding Wilson coefficients,
which contain the complete top- and $W$- mass dependence,
and $\lambda_t=V_{tb}V_{ts}^*$ with $V_{ij}$ being the
CKM matrix elements. The CKM dependence globally factorizes,
because we work in the approximation $\lambda_u=0$.

In the basis introduced by Misiak \cite{Mikolaj}, the operators read
\begin{eqnarray}
\label{operators}
O_1 &=& \left( \bar{s}_L \gamma^\mu T^a c_L \right) \,
        \left( \bar{c}_L \gamma_\mu T^a b_L \right) \,, \nonumber \\
O_2 &=& \left( \bar{s}_L \gamma^\mu  c_L \right) \,
        \left( \bar{c}_L \gamma_\mu  b_L \right) \,, \nonumber \\
O_7 &=& \frac{e}{16\pi^{2}} m_b(\mu) \bar{s} \, \sigma^{\mu \nu}
      \, R \, b \, F_{\mu \nu} \,,
        \nonumber \\
O_8 &=& \frac{g_s}{16\pi^{2}} m_b(\mu) \bar{s} \, \sigma^{\mu \nu}
      \, R \, T^a \, b \, G^a_{\mu \nu} \quad .
\end{eqnarray}
As the Wilson coefficients of the QCD penguin
operators $O_3,...,O_6$ are small, we do not list them here.

A consistent calculation for 
$b \to s \gamma$ at NLL precision requires three steps:
\begin{itemize}
\item[{\it 1)}] 
a matching calculation of the full standard model theory 
with the effective theory at the scale $\mu=\mu_W$ 
to order $\alpha_s^1$ for the Wilson coefficients, 
where  $\mu_W$ denotes a scale of order $M_W$ or $m_t$;
\item[{\it 2)}]  
a renormalization group evolution of the Wilson coefficients
from the matching scale $\mu_W$ down to the low scale
$\mu_b=\ordo{m_b}$,
using the anomalous-dimension matrix to order $\alpha_s^2$;
\item[{\it 3)}]   
a calculation of the matrix elements of the operators at the scale 
$\mu = \mu_b$  to order $\alpha_s^1$. 
\end{itemize}

As all three steps are rather involved, a common effort of several
independent groups was needed in order to calculate the NLL prediction
for ${\rm BR}(B \to X_s \gamma)$ 
\cite{counterterm,Adel,GH,INFRARED,Mikolaj,AG91,Pott,GHW}.
For a detailed summary of the various steps
and intermediate results, we refer to the recent review by T. Hurth 
\cite{Hurth_rev}. However,
we would like to point out that the most difficult part, viz.
the calculation of three-loop anomalous dimensions performed
by Chetyrkin, Misiak and M\"unz in 1996 \cite{Mikolaj}, was
only confirmed very recently by Gambino, Gorbahn and Haisch
\cite{Gambino:2003zm}. Their paper also contains the three-loop mixing
of the four-Fermi operators into $O_9$, which is important 
for the process $B \to X_s \ell^+ \ell^-$.

During the completion of the NLL QCD corrections, also calculations of 
electroweak corrections were started 
\cite{Czarnecki:1998tn,Kagan:1999ym,Baranowski:1999tq}. 
At present, the corrections 
of order $\alpha_{\rm em} \ln(\mu_b/M)$ $[\alpha_s \ln(\mu_b/M)]^n$, 
as well as the subleading terms
of order $\alpha_{\rm em}[\alpha_s \ln(\mu_b/M)]^n$ \cite{GH00}
are systematically available.

\section{NLL (and partial NNLL) results for {\bfseries BR}$\bm{(B \to X_s \gamma)}$}
Combining NLL QCD corrections with the electroweak corrections just
mentioned and also including the $1/m_b^2$ \cite{powermb}
and $1/m_c^2$ \cite{powermc} power corrections, 
the branching ratio reads
\begin{equation}
{\rm BR}(B \to X_s \gamma) = (3.32  \pm 0.14 \pm 0.26) \times 10^{-4}  \, , 
\label{NLLmcpole}
\end{equation}
where the first error reflects the dependence on the renormalization scale
$\mu_b$ varied in the interval
$m_b/2 \le \mu_b \le 2 m_b$, while the second error reflects the error
due to the uncertainties in the input parameters.  

Among the input parameters the charm quark mass $m_c$ plays a crucial role.
The charm quark mass dependence only enters the prediction for the
decay width at the NLL level, more precisely through the 
$\ordo{\alpha_s}$ correction to the matrix elements $\langle s \gamma|O_{1,2}|b
\rangle$. Until recently, all authors used the pole mass
value $m_c^{\rm{pole}}$ for the charm quark mass
in numerical evaluations, leading to a branching ratio as 
specified in eq. (\ref{NLLmcpole}). 

In 2001, however, Gambino and Misiak \cite{Gambino:2001ew} 
pointed out that the MS-bar
mass $\overline{m}_c$, normalized at $\mu \approx m_b/2$, could
be the better choice, because the charm quark appears as an off-shell particle
in the loop involved in the above mentioned matrix element with 
a typical virtuality of $m_b/2$. 
Using this interpretation, $\overline{m}_c/m_b=0.22\pm 0.04$
is substantially smaller than the value  
$m_c^{\rm{pole}}/m_b=0.29\pm 0.02$ used in eq. (\ref{NLLmcpole}), leading
to a branching ratio of \cite{Gambino:2001ew,Buras:2002tp}
\begin{equation}
 {\rm BR}(B \to X_s \gamma) = (3.70  \pm 0.30) \times 10^{-4}  \, . 
\label{NLLmcmsbar}
\end{equation}

We would like to stress here that the above argument in favour
of $\overline{m}_c$ is an intuitive one. 
Formally, the difference between using $m_c^{\rm{pole}}$ 
or $\overline{m}_c$
amounts to a NNLL effect at the level of the branching
ratio. This means that a NNLL becomes necessary in order to unambiguously fix 
this issue.

Before sketching the NNLL program, we would like to stress that
settling the $m_c$ issue is also important when extracting bounds
on new physics, based on NLL calculations \cite{Ciuchini:1997xe,BG98}. 
For example, in the type-II two-Higgs-doublet model,
one obtains a bound from $b \to s \gamma$ on the charged
Higgs boson mass of $m_H >350$ GeV ($99\%$ C.L.) when using 
$\overline{m}_c$. When using on the other hand $m_c^{\rm{pole}}$, the bound
is   $m_H >280$ GeV ($99\%$ C.L.) \cite{Gambino:2001ew}.

\begin{figure}[tb]
\centerline{
\resizebox{0.75\hsize}{!}{
\includegraphics*{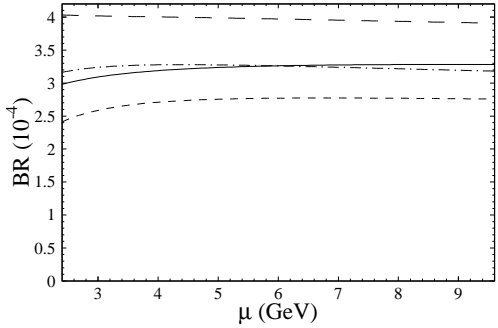}
}}
\caption{${\rm BR}(B \to X_s \gamma)$ as a function of the renormalization
scale $\mu$. The dash-dotted curve shows the NLL prediction; the solid
curve contains in addition the $\ordo{\alpha_s^2 n_f}$ terms.
The long-dashed (short-dashed) curve is obtained by switching off the 
$\ordo{\alpha_s^2 n_f}$ corrections to $O_7$ ($O_2$). Figure taken from ref. 
\cite{Bieri:2003ue}.}
\label{fig:2}
\end{figure}

Concerning the NNLL program, it is clear that in order to get a full NNLL QCD
result for ${\rm BR}(B \to X_s \gamma)$, all the three steps listed above have
to be improved by one order in $\alpha_s$.  This means that three-loop
matching calculations are needed, up to four-loop anomalous dimensions have to
be worked out and up to three-loop calculations at the level of the matrix
elements $\langle s \gamma|O_i(\mu_b)|b \rangle$ have to be performed.
Several groups have been formed in order to attack this ambitious goal.
Recently, a calculation of the $\ordo{\alpha_s^2 n_f}$ corrections to the
matrix elements of the operators $O_1$, $O_2$, $O_7$ and $O_8$ was published
\cite{Bieri:2003ue}.
Diagrammatically, these contributions are generated by inserting quark
bubbles ($n_f$ denotes the number of light quarks) 
into the gluon propagators in the diagrams which are involved
in the calculations at NLL order. We note that these contributions are 
not related to the definition problem of $m_c$. However, in many other
cases they are sources of large corrections. E.g., in the semileptonic decay
width $\Gamma(B \to X_c \ell \nu_\ell)$ these $\ordo{\alpha_s^2 n_f}$
terms (after replacing $n_f \to -3\beta_0/2$, according to the procedure of
naive non-abelianization) incorporate more than $80\%$ of the 
complete $\ordo{\alpha_s^2}$
corrections \cite{Czarnecki:1998kt}. 
The impact of the $\ordo{\alpha_s^2 n_f}$ corrections to
${\rm BR}(B \to X_s \gamma)$ are shown in fig. \ref{fig:2}.
The dash-dotted curve shows the NLL prediction, while
the solid
curve incorporates in addition the $\ordo{\alpha_s^2 n_f}$ terms (after the
replacement $n_f \to  -3 \beta_0/2$, according to naive non-abelianization).
As one sees from the figure, the $\ordo{\alpha_s^2 n_f}$ corrections seem
to be small. Note, however, that this is a result of a relatively large
accidental cancellation between corrections to $O_2$ and $O_7$. This
point is illustrated by the long-dashed and short-dashed curve, which are 
obtained by switching off the 
$\ordo{\alpha_s^2 n_f}$ corrections to $O_7$ and $O_2$, respectively.
 
\section{Partially integrated BR and photon energy spectrum}
The photon energy spectrum of the partonic decay $b \to s \gamma$
is a delta function, concentrated at $\sim (m_b/2)$, 
when the $b$-quark decays at rest. This delta function gets smeared when
considering the inclusive photon energy spectrum
from a $B$ meson decay. There is a perturbative contribution to this smearing, 
induced by
the Bremsstrahlung process $b \to s \gamma g$ \cite{AG91,Pott},
as well as a nonperturbative
one, which is due to the Fermi motion of the decaying $b$ quark in 
the $B$ meson. 

For small photon energies, the $\gamma$-spectrum from $B \to X_s \gamma$ is
completely overshadowed by background processes, 
like $b \to c \bar{u} d \gamma$ and $b \to u \bar{u} d \gamma$.
This background falls off very rapidly with increasing photon
energy, and becomes small for $E_\gamma > 2$ GeV \cite{AG92}.
This implies
that only the partial branching ratio
\begin{equation}
{\rm BR }(B \to X_s \gamma)(E_\gamma^{\rm{min}}) = 
\int_{E_\gamma^{\rm{min}}}^{E_\gamma^{\rm{max}}}
 \frac{d{\rm BR}}{dE_\gamma}
dE_\gamma
\end{equation}
can be directly measured, with $E_\gamma^{\rm{min}}=\ordo{2}$ GeV.

Putting the energy cut at $E_\gamma^{\rm{min}} = 2.0$ GeV, 
CLEO used two methods to analyze
their data on the photon energy spectrum in their most recent analysis: 
First, the Ali-Greub model \cite{AG91,AGmodel}, 
based on the spectator model formulated
in ref. \cite{ACCMM} and second, methods based on HQET \cite{Kagan:1999ym}.
  
\begin{figure}[tb]
\centerline{
\resizebox{0.55\hsize}{!}{
\includegraphics*{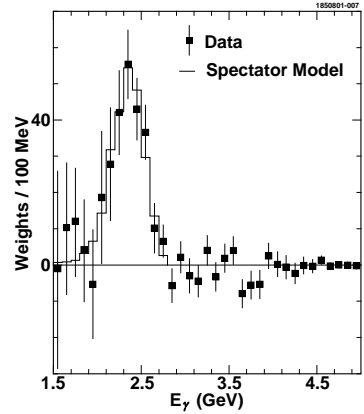}
}}
\caption{Photon energy spectrum: The data points represent the
CLEO measurement \cite{CLEOnew}. 
The histogram shows the theory result based on the \
spector model using $p_F=410$ MeV and $\langle m_b \rangle=4.69$ GeV.
Fig. taken from ref. \cite{CLEOnew}.}
\label{fig:3}
\end{figure}

The spectator model contains two free parameters, viz. $p_F$, the average
Fermi momentum of the $b$ quark in the $B$ meson and the mass of the spectator 
quark, $m_{\rm{spec}}$.
Equivalently $(p_F,\langle m_b \rangle)$ can be used as the free parameters, 
where $\langle m_b \rangle$ is the average $b$ quark mass as defined in ref. 
\cite{AG91,AGmodel}. In Fig. \ref{fig:3} a comparison between theory and experiment
is shown. Using $p_F=410$ MeV and $\langle m_b \rangle=4.69$ GeV the best
fit is obtained. We would like to stress that similar values for these
parameters are also obtained when fitting the lepton spectra in 
$B \to X_c \ell \nu$ and $B \to X_u \ell \nu$.

A modern way - based on first principles -  implements 
the Fermi motion in the framework
of the heavy-quark expansion. When probing the spectrum
closer to the endpoint, the OPE breaks down, and the leading twist
nonperturbative corrections must be resummed into the $B$ meson structure
function $f(k_+)$ \cite{shape}, where $k_+$ is the
light-cone momentum of the $b$ quark in the $B$ meson.
The physical spectrum is then obtained by
the convolution 
\begin{equation}
\frac{d\Gamma}{dE_\gamma} = \int_{2E_\gamma-m_b}^{\bar{\Lambda}} dk_+ f(k_+)
\frac{d\Gamma_{{\rm part}}}{dE_\gamma}(m_b^*) \quad ,
\end{equation}
where $(d\Gamma_{\rm part}/dE_\gamma)(m_b^*)$ 
is the partonic
differential rate, written as a function of the ``effective mass''
$m_b^*=m_b+k_+$. 
The function $f(k_+)$ has support in the range $-\infty < k_+ <
\bar{\Lambda}$, where $\bar{\Lambda}=m_B-m_b$ 
in the infinite mass limit. This implies that
the addition of the structure function 
moves the partonic endpoint
of the spectrum from $m_b/2$ to the physical endpoint $m_B/2$.
While the shape of the function $f(k_+)$ is unknown, the first few
moments $A_n = \int dk_+ \, k_+^n f(k_+)$ are known: $A_0=1$, $A_1=0$
and $A_2 = -\lambda_1/3$. 
As $A_n$ ($n>2$) are poorly known, several Ans\"atze
were used for $f(k_+)$; e.g. Neubert and Kagan \cite{Kagan:1999ym} 
used $f(k_+)= N (1-x)^a
e^{(1+a)x}$, with $x=k_+/\bar{\Lambda}$. Taking into account the constraints 
from $A_0$, $A_1$ and $A_2$, the independent parameters in this Ansatz
can be chosen to be $m_b$ and $\lambda_1$. 
As shown in \cite{Kagan:1999ym}, the uncertainty of 
$m_b$ dominates the error of the partial branching ratio. 
\begin{figure}[tb]
\centerline{
\resizebox{0.65\hsize}{!}{
\includegraphics*{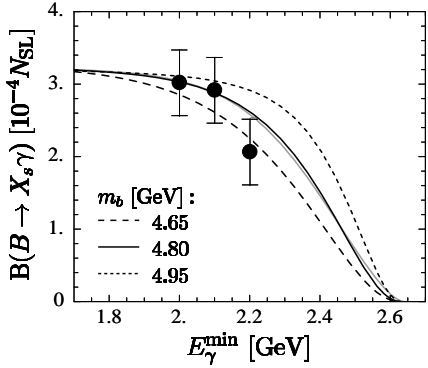}
}}
\caption{Partially integrated branching ratio as a function of
         the energy cutoff $E_\gamma^{\rm{min}}$; 
         Curves taken from Kagan and Neubert \cite{Kagan:1999ym}.
         Data point represent CLEO measurements 
         \cite{CLEOold,CLEOmed,CLEOnew}.}
        \label{fig:neubert}
\end{figure}
In figure~\ref{fig:neubert}
the partial branching ratio is shown for the relevant range of $m_b$
as a function of $E_\gamma^{\rm{min}}$, keeping $\lambda_1/\bar{\Lambda}^2$ fixed.
The data points show three CLEO measurements. In the oldest one the photon
energy cut was put at $E_\gamma^{\rm{min}}=2.2$ GeV, while in the most recent
analysis this cut was lowered to 2.0 GeV, which is very important, because
at 2.0 GeV the theoretical error on the partial branching ratio is
considerably smaller, as seen from fig.~\ref{fig:neubert}.    

To determine from the measurement of the partial branching ratio the full BR,
one needs from theory the fraction $R$ of the $B \to X_s \gamma$
events with photon energies above $E_\gamma^{\rm{min}}$. Based on Kagan-Neubert
\cite{Kagan:1999ym},
CLEO \cite{CLEOnew} obtained $R=\left( 0.915 ^{+0.027}_{-0.055} \right)$. 
A similar result is also obtained when using the spectator model.

It has been shown that up to corrections of $\ordo{\Lambda_{\rm QCD}/m_b}$,
the same shape function also describes $B \to X_u \ell \nu$
\cite{Neubert:1993um}.  This implies that the photon energy spectrum can be
used to predict the fraction of $B \to X_u \ell \nu$ events with
$E_{\rm{lept}} > 2.2$ GeV, where leptons coming from $B \to X_c \ell \nu$ are
absent for kinematical reasons.  Taking into account perturbative and
$\Lambda_{\rm{QCD}}/m_b$ corrections
\cite{Leibovich:1999xf,Bauer:2002yu,Neubert:2002yx}, it is possible to extract
$V_{ub}$ from a measurement of the $B \to X_u \ell \nu$ decay rate in the
region above 2.2 GeV.  CLEO used this strategy in ref. \cite{CLEO2002} to
extract the CKM matrix element $|V_{ub}|$, obtaining $|V_{ub}|=(4.08 \pm
0.56_{\rm{exp}} \pm 0.29_{\rm{th}})\times 10^{-3}$.

\section{$\bm{B \to X_d \gamma}$ in the SM}
The decay $B \to X_d \gamma$ can be treated in a similar way as
$B \to X_s \gamma$ \cite{AAG98}.
The only difference is that
$\lambda_u$ for $b \to d \gamma$ is not small relative to
$\lambda_t$ and $\lambda_c$; therefore, also
the current-current operators $O_1^u$ and $O_2^u$, 
weighted by $\lambda_u$, contribute.
\begin{figure}[H]
\centerline{
\resizebox{0.45\hsize}{!}{
\includegraphics*{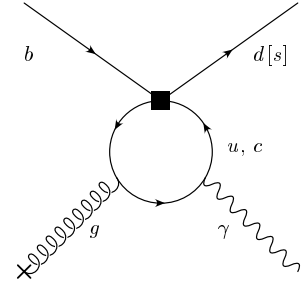}
}}
\caption{Interaction of the $c-$ and $u-$quark loop with soft gluons
surrounding the $b$ quark in the $B$ meson.}
        \label{fig:5}
\end{figure}
Unfortunately, these operators induce long-distance contributions to $B \to
X_d \gamma$, which at present are not very well understood.  To illustrate the
problem, we first look at the corresponding charm quark loop, depicted in
fig. \ref{fig:5}.  In this case, one can expand the loop function
\[
\sim \int_0^1 dx \int_0^{1-x} dy 
\frac{xy}{m_c^2\left[ 1- \frac{k_g^2}{m_c^2}x(1-x)-
2 x y \frac{k_g k_\gamma}{m_c^2} \right]}
\]
in powers of $t=k_g k_\gamma/m_c^2$, where $k_g$ and $k_\gamma$ denote
the momentum of the gluon and the photon, respectively.
This expansion generates the so-called
Voloshin terms \cite{powermc}, which in ${\rm BR}(B \to X_s \gamma)$ are 
a $3\%$ effect.
Obviously, there is no such OPE in the case of the $u-$quark loop. However,
Buchalla, Isidori and Rey \cite{BIR} argued that an expansion in $1/t$ can be
done, leading to non-local operators. From naive dimensional counting, the
leading contribution is expected to be of order $\Lambda_{\rm{QCD}}/m_b$. 

In reference~\cite{AAG98}, where NLL calculations for the process
$B \to X_d \gamma$ were presented,
the uncertainties due to the long-distance effects were absorbed into
the theoretical error.
Using $\mu_b=2.5$ GeV and the central values of the input
parameters, the analysis in reference~\cite{AAG98} yields 
a difference between the LL and NLL predictions
for ${\rm BR}(B \to X_d \gamma) $ of 
$\sim 10\%$, increasing the branching ratio in the NLL case.
For a fixed value of the CKM-Wolfenstein parameters 
$\rho$ and $\eta$, the theoretical uncertainty of the average branching
ratio $\langle {\rm BR}(B \to X_d \gamma) \rangle$ of the decay $B \to X_d \gamma$
and its charge conjugate $\overline{B} \to \overline{X_d} \gamma$ is:
$\Delta \langle {\rm BR}(B \rightarrow X_d \gamma)\rangle/ 
\langle {\rm BR}(B \rightarrow X_d \gamma) \rangle   
= 
\pm (6-10)\%$. 
Of particular theoretical interest for constraining $\rho$ and $\eta$
is the ratio of the
branching ratios, defined as
\begin{equation}
\label{dsgamma}
R(d\gamma/s\gamma) \equiv \frac{\langle {\rm BR}(B \to X_d \gamma) \rangle}
                           {\langle {\rm BR}(B \to X_s \gamma) \rangle},
\end{equation}
in which a good part of the theoretical uncertainties cancels. 
Varying the CKM-Wolfenstein parameters $\rho$ and $\eta$ in the range
$-0.1 \leq \rho \leq 0.4$ and $0.2 \leq \eta \leq 0.46$ and taking into
account other parametric dependences, the 
results (without electroweak corrections) are
\begin{eqnarray}
\label{summarybrasy}
6.0 \times 10^{-6} &\leq &
{\rm BR}(B \rightarrow X_d \gamma)   \leq 2.6 \times 10^{-5}~, \nonumber\\
0.017 &\leq & R(d\gamma/s\gamma) \leq 0.074~.\nonumber
\end{eqnarray}
Another observable, which is also sensitive to the CKM parameters
$\rho$ and $\eta$, is the CP rate asymmetry $a_{{\rm CP}}$, defined as
\begin{equation} 
a_{{\rm CP}} = 
\frac{\Gamma(B \to X_d \gamma)-\Gamma(\overline{B} \to \overline{X_d} \gamma)}{
      \Gamma(B \to X_d \gamma)+\Gamma(\overline{B} \to \overline{X_d} \gamma)}
    \, .
\end{equation} 
Varying $\rho$ and $\eta$ in the range specified above, one gets
$7\% \le a_{{\rm CP}} \le 35\%$ \cite{AAG98}.
We would like to point out that $a_{{\rm CP}}$ is at the moment only available
to LL precision and therefore suffers from a relatively large renormalization
scale dependence.

In summary, this decay mode is very challenging, both in theory and
experiment:
On the theory side more work is needed concerning the nonperturbative 
contributions associated with the $u-$quark loop,  while on the experimental
side the observation of this decay needs high statistics and a very good
discrimination between pions and kaons.

\begin{acknowledgement}

This work is partially supported by the Swiss National
Foundation and by RTN, BBW-Contract N0. 01.0357 and EC-Contract
HPRN-CT-2002-00311 (EURIDICE).

\end{acknowledgement}


\end{document}